\documentclass[twocolumn]{aastex631}
\usepackage{textgreek}
\usepackage{needspace, longtable, placeins, bm}
\usepackage{gensymb}
\usepackage{textgreek}

\received{January 5, 2023}
\revised{March 14, 2023}
\accepted{April 24, 2023}

\submitjournal{AJ}

\shorttitle{Mid-to-Late M-Dwarfs Lack Jupiter Analogs}
\shortauthors{Pass et al.}
\graphicspath{{./}{figures/}}

\begin{document}

\title{Mid-to-Late M Dwarfs Lack Jupiter Analogs}

\widowpenalty=10

\author[0000-0002-1533-9029]{Emily K. Pass}
\affiliation{Center for Astrophysics $\vert$ Harvard \& Smithsonian, 60 Garden Street, Cambridge, MA 02138, USA}

\author[0000-0001-6031-9513]{Jennifer G. Winters}
\affiliation{Center for Astrophysics $\vert$ Harvard \& Smithsonian, 60 Garden Street, Cambridge, MA 02138, USA}
\affiliation{Thompson Physics Lab, Williams College, 880 Main Street, Williamstown, MA 01267, USA}

\author[0000-0002-9003-484X]{David Charbonneau}
\affiliation{Center for Astrophysics $\vert$ Harvard \& Smithsonian, 60 Garden Street, Cambridge, MA 02138, USA}

\author{Jonathan M. Irwin}
\affiliation{Center for Astrophysics $\vert$ Harvard \& Smithsonian, 60 Garden Street, Cambridge, MA 02138, USA}
\affiliation{Institute of Astronomy, University of Cambridge, Madingley Road, Cambridge CB3 0HA, UK}

\author[0000-0001-9911-7388]{David W. Latham}
\affiliation{Center for Astrophysics $\vert$ Harvard \& Smithsonian, 60 Garden Street, Cambridge, MA 02138, USA}

\author{Perry Berlind}
\affiliation{Center for Astrophysics $\vert$ Harvard \& Smithsonian, 60 Garden Street, Cambridge, MA 02138, USA}

\author[0000-0002-2830-5661]{Michael L. Calkins}
\affiliation{Center for Astrophysics $\vert$ Harvard \& Smithsonian, 60 Garden Street, Cambridge, MA 02138, USA}

\author[0000-0002-9789-5474]{Gilbert A. Esquerdo}
\affiliation{Center for Astrophysics $\vert$ Harvard \& Smithsonian, 60 Garden Street, Cambridge, MA 02138, USA}

\author[0000-0003-3594-1823]{Jessica Mink}
\affiliation{Center for Astrophysics $\vert$ Harvard \& Smithsonian, 60 Garden Street, Cambridge, MA 02138, USA}



\begin{abstract}

\noindent Cold Jovian planets play an important role in sculpting the dynamical environment in which inner terrestrial planets form. The core accretion model predicts that giant planets cannot form around low-mass M dwarfs, although this idea has been challenged by recent planet discoveries. Here, we investigate the occurrence rate of giant planets around low-mass (0.1-0.3M$_\odot$) M dwarfs. We monitor a volume-complete, inactive sample of 200 such stars located within 15 parsecs, collecting four high-resolution spectra of each M dwarf over six years and performing intensive follow-up monitoring of two candidate radial-velocity variables. We use TRES on the 1.5 m telescope at the Fred Lawrence Whipple Observatory and CHIRON on the Cerro Tololo Inter-American Observatory 1.5 m telescope for our primary campaign, and MAROON-X on Gemini North for high-precision follow-up. We place a 95\%-confidence upper limit of 1.5\% (68\%-confidence limit of 0.57\%) on the occurrence of $M_{\rm P}$sin$i > $1M$_{\rm J}$ giant planets out to the water snow line and provide additional constraints on the giant planet population as a function of $M_{\rm P}$sin$i$ and period. Beyond the snow line ($100$ K $< T_{\rm eq} < 150$ K), we place 95\%-confidence upper limits of 1.5\%, 1.7\%, and 4.4\% (68\%-confidence limits of 0.58\%, 0.66\%, and 1.7\%) for 3M$_{\rm J} < M_{\rm P}$sin$i < 10$M$_{\rm J}$, 0.8M$_{\rm J} < M_{\rm P}$sin$i < 3$M$_{\rm J}$, and 0.3M$_{\rm J} < M_{\rm P}$sin$i < 0.8$M$_{\rm J}$ giant planets; i.e., Jupiter analogs are rare around low-mass M dwarfs. In contrast, surveys of Sun-like stars have found that their giant planets are most common at these Jupiter-like instellations.
\end{abstract}

\section{Introduction} \label{sec:intro}
Recent work has shown that Jupiter-mass planets at Jovian instellations are a common occurrence around Sun-like stars, and the population peaks just beyond the snow line \citep{Fulton2021, Rosenthal2022}. But is the same true around much less massive stars?  The core accretion theory predicts that giant planets are rare around M dwarfs, as giant planet formation within the gas dispersal time is inhibited by increasing dynamical time and decreasing mass surface density with decreasing stellar mass \citep{Laughlin2004}. In accordance with this prediction, only a handful of M-dwarf giant planets have been identified by radial-velocity \citep{Marcy1998, Marcy2001, Butler2006, Johnson2007, Johnson2010, Bailey2008, Apps2010, Haghighipour2010, Howard2010, Morales2019, Endl2022} and transit investigations \citep{Johnson2012, Hartman2015, Bayliss2017, Bakos2020, Canas2020, Gan2021, Jordan2021, Canas2022, Kanodia2022, Kanodia2023}. Other such planets have been inferred through microlensing events \citep[e.g.,][]{Kains2013, Mroz2017}. However, individual detections are only anecdotal evidence. To measure occurrence rates, one must turn to surveys for which the sensitivity and completeness have been carefully investigated.

Various works have studied the occurrence rate of giant planets around M dwarfs, with each work reporting these rates in differing bins. \citet{Bonfils_2013} determine an occurrence rate of $<1$\% for $M_{\rm P}$sin$i = 100-1000$M$_\oplus$ planets with periods of 100--1000 days and 4$^{+5}_{-1}$\% for periods of 1000--10000 days. \citet{Montet2014} report a 6.5$\pm$3.0\% occurrence rate for $1M_{\rm J} < M_{\rm P} < 13M_{\rm J}$ planets out to 20 au. \citet{Sabotta2021} find an occurrence rate of 6$^{+4}_{-3}$\% for M$_{\rm P}$sin$i = 100-1000$M$_\oplus$ planets with periods of up to 1000 days. Combining the \citet{Bonfils_2013} radial-velocity and \citet{Gould2010} microlensing results, \citet{Clanton2014} measure an occurrence rate of 2.9$^{+1.3}_{-1.5}$\% for 1M$_{\rm J} < M_{\rm P}$sin$i < 13$M$_{\rm J}$ planets with \hbox{$1 < P\rm < 10^4$-day} periods around M dwarfs, making Jupiter-size planets 4.3 times rarer around M dwarfs than around FGK stars. However, the definition of M dwarfs is broad, spanning nearly an order of magnitude in stellar mass \citep[0.08--0.62M$_\odot$;][]{Benedict2016}. Extant occurrence rate constraints are based on samples dominated by the higher-mass end; for example, \citet{Gould2010} report that their lens distribution is centered at $M_*=0.5$M$_\odot$, while the mean stellar mass in the \citet{Bonfils_2013} sample is 0.35M$_\odot$. Given the stellar mass dependence of the factors opposing giant planet formation around M dwarfs, low-mass M dwarfs are best suited to probing tensions with the core formation theory.

Only two of the known M-dwarf giant planets orbit $M_{*} \leq 0.3$M$_\odot$ stars: LHS 252 b, a 0.46M$_{\rm J}$ planet with a 204-day period \citep[][]{Morales2019}, and GJ 83.1 b, a 0.21M$_{\rm J}$ planet with a 771-day period \citep{Feng2020, Quirrenbach2022}. An additional candidate has been proposed by \citet[][]{Curiel2020}: TVLM 513-46 b, a 0.4M$_{\rm J}$ planet with a 221-day period. 2MA J1227-7227 b, a young 0.85R$_{\rm J}$ planet with a 27.4-day period, may be an additional such object, although \citet{Mann2021} suggest that it will eventually contract into a sub-Neptune. The directly imaged planet 2MA J0437+2651 b has also been interpreted as a super-Jupiter around a young 0.15-0.18M$_\odot$ star \citep{Gaidos2021}. Despite this small sample, tension with core accretion theory is already emerging, as \citet{Morales2019} find that the properties of LHS 252 b cannot be reproduced through existing core accretion models. The authors suggest formation via gravitational instability \citep{Boss2006} as a possible alternative. These tensions are further discussed in \citet{Schlecker2022}.

Previous radial-velocity studies investigating the occurrence rate of giant planets around M dwarfs have included too few low-mass M dwarfs to robustly constrain the occurrence rate of giant planets in this population (Figure 1; for consistency in comparing between studies, we calculate all masses in this figure using Gaia parallaxes and the $K$-band mass-luminosity relation from \citealt{Benedict2016}). Using a small sample of 23 M dwarfs less massive than 0.34M$_\odot$, \citet{Sabotta2021} estimate an occurrence rate of 16\%$^{+15\%}_{-9\%}$ for planets with M$_{\rm P}$sin$i = 100-1000$M$_\oplus$ at periods of 100--1000 days, which could indicate an enhanced rate of giant planet formation around low-mass M dwarfs; however, they note that selection biases in this sample may result in this value being overestimated by up to a factor of five. A large, statistically robust sample is necessary to establish whether planets like LHS 252 b are rare oddities, or whether the core-accretion framework is fundamentally inaccurate for low-mass M dwarfs.

In addition to informing planet formation theories, the occurrence rate of giant planets is relevant to questions of the formation of terrestrial planets, their accretion of material from beyond the snow line, and ultimately their habitability. Due to their strong gravities, giant planets set the dynamical and collisional conditions under which terrestrial planets form \citep{Lunine2001}. A cold Jupiter may be necessary to create a planetary system like our own, playing a role in preventing the inward migration of gas giant cores \citep{Izidoro2015, Kruijer2017} and setting the water budget of the terrestrial planets \citep{Raymond2017, Bitsch2021}. As nearby M dwarfs represent the most favorable hosts for the detection and characterization of Earth-like planets with present and near-future instrumentation \citep{Charbonneau2007, Snellen2013, Morley2017, Lopez-Morales2019}, the occurrence of giant planets around the lowest-mass stars is relevant for evaluating the habitability of the systems that can be studied in detail in the foreseeable future.

In Section~\ref{sec:obs}, we present our spectroscopic survey of nearby 0.10--0.30M$_\odot$ M dwarfs. Section~\ref{sec:methods} describes our data reduction process, radial-velocity calculations, and error estimates. In Section~\ref{sec:analysis}, we discuss our identification and follow-up of radial-velocity variables and our calculation of occurrence rate constraints. We present discussion and comparison to the literature in Section~\ref{sec:discussion} and a summary in Section~\ref{sec:summary}.

\begin{figure}
    \centering
    \includegraphics[width=\columnwidth]{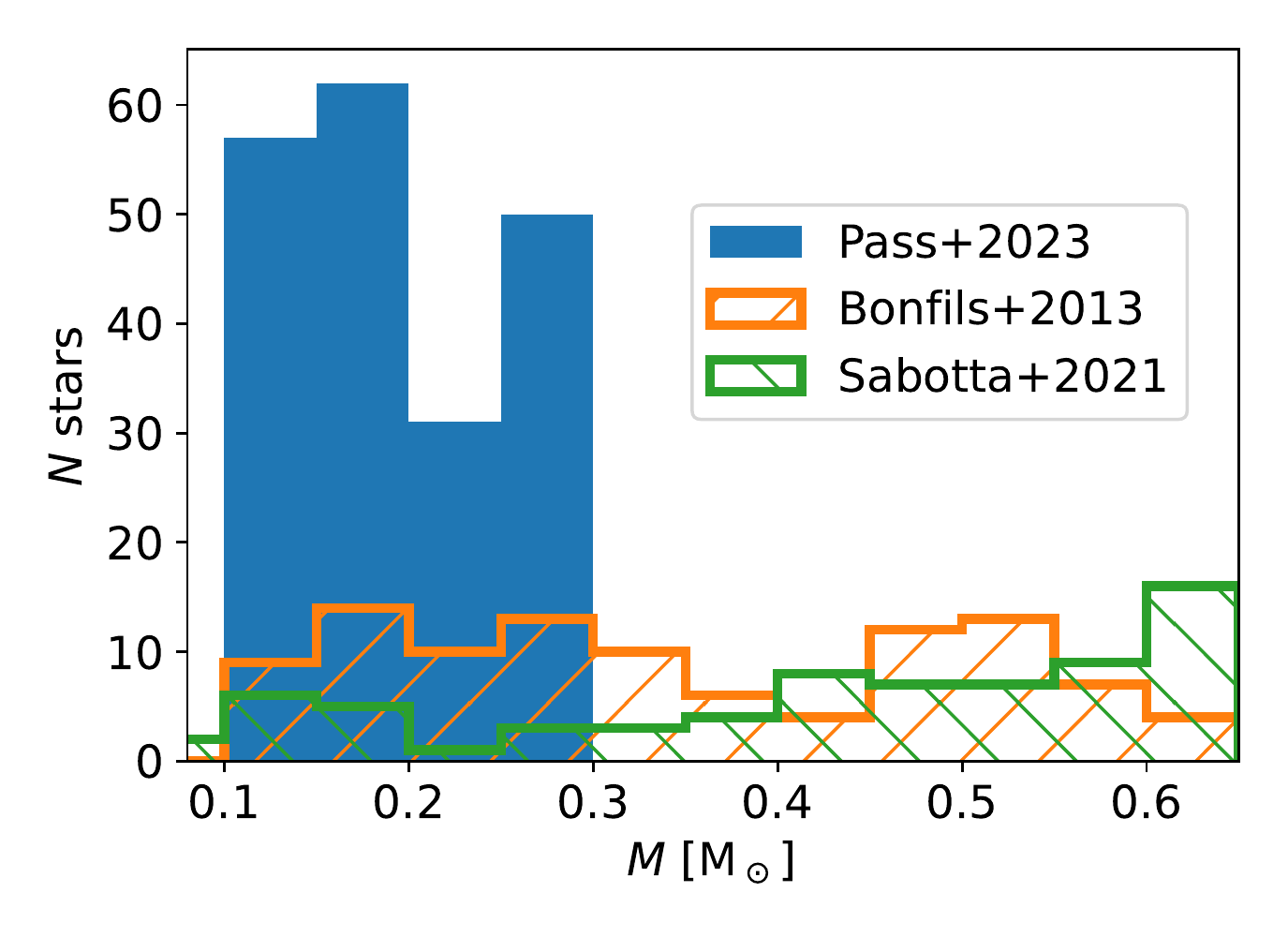}
    \vspace{-0.8cm}
    \caption{The mass distribution of our stars. We contrast this sample with \citet[][B13]{Bonfils_2013}, the largest radial-velocity survey of M dwarfs in the literature, as well as \citet[][S21]{Sabotta2021}, recent results from the CARMENES survey. Previous studies of giant planets around M dwarfs have been dominated by early Ms, with masses that are not so dissimilar to those of Sun-like stars. More poorly studied are low-mass, fully convective M dwarfs. Our sample is volume complete and contains 200 such stars with masses of 0.1--0.3M$_\odot$. Previous studies have dramatically smaller samples over this mass range (46 in B13 and 15 in S21) and are not volume complete. Note that this figure compares total sample size; for a comparison of effective sample size as a function of planetary mass and period, an interested reader may compare our Figure~\ref{fig:occ} with Figure 15 of B13 and Figure 3 of S21.}
    \label{fig:samp}
\end{figure}

\section{Stellar Sample and Data Collection}
\label{sec:obs}
This work is part of a series of papers presenting the results of the volume-complete spectroscopic survey of 0.10--0.30M$_\odot$ M dwarfs within \hbox{15 pc}, defined in \citet{Winters2021}. The sample totals 413 stars and excludes M dwarfs that are close companions (separations $<4$") to more massive primaries, as the spectra would be strongly contaminated by the brighter star. Between 2016 and 2022, we collected four high-resolution spectra of each of these 413 stars using the Tillinghast Reflector Echelle Spectrograph (TRES; $R=44000$) at the \hbox{1.5 m} telescope at the Fred Lawrence Whipple Observatory for sources with \hbox{$\delta > -15\degree$} and the CTIO HIgh ResolutiON (CHIRON; $R=80000$) spectrograph at the Cerro Tololo Inter-American Observatory \hbox{1.5 m} telescope for sources with \hbox{$\delta < -15\degree$.} Exposure times were selected based on a target per-pixel SNR of 15 in the TiO bands around 7100\AA, ranging from 60 to 5400 seconds; however, in practice our observations span a range of SNR, with a typical value of 11 for our TRES observations and 8 for our CHIRON observations. Forthcoming entries in this series will discuss the active subsample, the binary subsample, and present galactic kinematics for all stars (Pass et al.\ in press; Winters et al.\ in prep).

In this work, we consider only the single, inactive subsample, as with four observations per star, we cannot easily distinguish variation due to an orbiting planet from other sources of variation. As we perform intensive follow-up vetting of each candidate flagged as potentially variable from our four-observation campaign, selecting the single, inactive subsample allows us to minimize the time spent intensively monitoring candidates that are ultimately false positives. We discard all binaries separated by less than 4", as light from both stars would fall in the spectroscopic aperture under typical seeing. To avoid a large number of false positives due to activity-induced radial-velocity variability \citep[e.g.,][]{Tal-Or2018}, we neglect active stars for which we measure H\textalpha\ emission stronger than a median equivalent width (EW) of \hbox{-1\AA} using the method of \cite{Medina2020}. We use the notation that a negative EW indicates emission. This \hbox{-1\AA} threshold has been used to distinguish between active and inactive M dwarfs in previous work such as \citet{Newton2017}. Of the stars without close binaries, we classify 123 as active based on this criterion; this group includes the known giant planet hosts GJ 83.1 (H\textalpha\ EW$=-2.3$\AA) and LHS 252 (H\textalpha\ EW$=-1.8$\AA). While `inactive' M dwarfs may still have activity-induced variability that masquerades as a planet \citep[e.g.,][]{Lubin2021}, this phenomenon is less ubiquitous and can be evaluated on a case-by-case basis through intensive follow-up of the small number of inactive candidates. After making these cuts, 200 M dwarfs remain in our sample (Figure~\ref{fig:samp}), of which we observe 122 with TRES and 78 with CHIRON.

\section{Data Reduction}
\label{sec:methods}
\subsection{Radial velocities} \label{sec:RVs}
We extract the spectra using the standard TRES \citep{Buchhave2010} and CHIRON \citep{Tokovinin2013} pipelines and measure radial velocities using an updated version of the method presented in \cite{Winters2020}, which itself is based on \cite{Kurtz1998}. In addition to utilizing the TiO bandhead features from 7065--7165\AA\ (TRES echelle order 41 and CHIRON echelle order 44), we also consider five additional red echelle orders -- 36, 38, 39, 43, and 45 for TRES and 36, 37, 39, 40, and 51 for CHIRON, which represent wavelengths with low telluric contamination and high information content for low-mass M dwarfs. These orders fall within 6400--7850\AA.

We obtain an initial radial-velocity estimate for each spectrum by cross-correlating with a template of Barnard's Star in the manner described in \cite{Winters2020}. This method also produces an estimate of $v$sin$i$. We then shift and stack our observations to create a high-SNR template spectrum of the average slowly rotating, low-mass M dwarf observed by TRES. To stack the observations, we first create an error-weighted-mean spectrum for each star. We then median combine the spectra of all 122 stars that were observed with TRES. Prior to coaddition, we mask regions with telluric depths greater than 2\% in a representative TAPAS spectrum \citep{Bertaux2014} or skyline emission greater than $5\times10^{-17}$ erg/s/cm$^2$/\AA\ in the UVES sky emission atlas \citep{Hanuschik2003}. Our final template neglects wavelengths at the edges of orders that do not include information from at least ten M dwarfs.

In creating this average low-mass M dwarf template, we note that the information content $Q$ \citep{Bouchy2001} in TRES echelle order 45 differs by a factor of two across our sample, with the highest values obtained for cold, metal-rich stars and the lowest values obtained for hot, metal-poor stars. To minimize radial-velocity uncertainties caused by template mismatch, we therefore create six different average templates for TRES. Each star is initially classified by the $Q$ we measure in order 45, dividing the sample equally into six preliminary bins. We then reclassify each star by determining the preliminary template that maximizes the cross-correlation coefficient, recreating the templates based on these refined classifications, and repeating until the classifications converge. This iterative process prevents stars from being misclassified in cases where their $Q$ is not representative of their spectral type (for example, when noise leads to inflated estimates of $Q$). We perform a similar analysis for CHIRON, although we create only three templates so that we attain an acceptable SNR given the smaller number of observations available for coaddition.

We compute the cross-correlation function as the weighted sum:

\begin{equation}
r_j = \frac{\sum_{i} {a_{i+j} b_i w_{i+j}}}{\sqrt{(\sum_{i} {a^2_{i+j} w_{i+j}}) (\sum_{i} {b^2_{i} w_{i+j}})}},
\end{equation}

\noindent where $a$ is the observed flux, $b$ is the template flux, and $w$ are the variance weights (or $w_i$ = 0 for masked telluric or skyline features). This sum is evaluated jointly across the six echelle orders. Both $a$ and $b$ are blaze-corrected, normalized, continuum-subtracted, and oversampled by a factor of 32. Our code also includes the ability to rotationally broaden $b$ to analyze rapidly rotating stars, but there are no such stars in this study due to our H\textalpha-activity cut. The errors used to calculate $w$ consider photon noise and read noise in $a$, and are scaled based on the blaze correction.

Lastly, we determine the radial-velocity shift by fitting a Lorentzian to the cross-correlation peak.
 
\subsection{Uncertainties}
We calculate theoretical radial-velocity uncertainties using the equation derived in \cite{Bouchy2001}:

\begin{equation}
    \frac{\delta_{\rm RV}}{c} = \frac{1}{\sqrt{\sum_{i} \lambda_i^2 (\partial b_i / \partial \lambda_i)^2 w_i}},
\end{equation}

\noindent where $b$ is once again the template flux and $w$ are the variance weights. In the photon-noise limit, this equation simplifies to $\delta_{\rm RV} = c / (Q \times$SNR).

Alternatively, we can estimate the radial-velocity uncertainty directly from the cross-correlation function using the equation from \cite{Zucker2003}:

\begin{equation}
    \frac{\delta_{\rm RV}}{c} = \sqrt{- \left( N_{\rm eff} \frac{r''_{\rm max}}{r_{\rm max}} \frac{r^2_{\rm max}}{1-r^2_{\rm max}}\right)^{-1}},
\end{equation}

\noindent where we have replaced $N$ with $N_{\rm eff} \equiv (\sum_{i} {w_i})^2 / \sum_{i} {w_i^2}$ to account for our error weighting. In this equation, $r_{\rm max}$ denotes the cross-correlation peak and $r''_{\rm max}$ represents its second derivative.

Generally, we find good agreement between these two uncertainty estimators, although they differ when template mismatch is a significant source of uncertainty (e.g., in the high-SNR limit or when there is substantial rotational broadening, although the latter is not relevant for the sample discussed here). We make the conservative choice to always adopt the larger of the two.

To validate our uncertainties, we also compute the cross-correlation function for each of the six echelle orders individually and compare the resultant radial velocities using the chi-squared estimator:

\begin{equation}
    \chi^2 = \sum_{k=1}^{6} {w_k (v_{\mathrm{rad}, k} - v_{\rm rad})^2},
\end{equation}

\noindent where the variance weights ($w_k = 1/\delta^2_k$) are determined using the above uncertainty calculations for each order. For both TRES and CHIRON, the resulting distribution of $\chi^2$ across all observations shows good agreement with a chi-squared distribution with five degrees of freedom, as we would expect for properly estimated uncertainties.

An independent source of uncertainty comes from the barycentric correction, which ideally would be calculated at the photon-weighted midpoint of the observation. As we lack information on the temporal flux distribution over an exposure, we adopt the geometric midpoint as the observation time. This approximation can produce a systematic error of up to 2 ms$^{-1}$ per minute of difference between the geometric and photon-weighted midpoints, as shown in \citet[][see their equation 8]{Tronsgaard2019}. That work found that the offset between these midpoints is typically around 5\% of the exposure length. For completeness, we therefore add a radial-velocity uncertainty term in quadrature corresponding to a 5\% offset to the midpoint time; however, we find that this contribution is ultimately negligible in the total error budget.

\begin{table*}[!t]
\caption{First five lines of the machine-readable data file listing our radial-velocity measurements and their uncertainties}
\label{tab:rvs}
\resizebox{1.\textwidth}{!}{ \hspace*{-3.5cm}\begin{tabular}{llrrrrrrrrrrrrl}
\hline
\multicolumn{1}{c}{Name} & \multicolumn{1}{c}{Inst} & \multicolumn{4}{c}{BJD - 2457640 [day]} & \multicolumn{4}{c}{Relative RV [ms$^{-1}$]} & \multicolumn{4}{c}{RV error [ms$^{-1}$]} & \multicolumn{1}{c}{$P(\chi^2)$} \\ \hline
 &  & \multicolumn{1}{c}{\textbf{1}} & \multicolumn{1}{c}{\textbf{2}} & \multicolumn{1}{c}{\textbf{3}} & \multicolumn{1}{c}{\textbf{4}} & \multicolumn{1}{c}{\textbf{1}} & \multicolumn{1}{c}{\textbf{2}} & \multicolumn{1}{c}{\textbf{3}} & \multicolumn{1}{c}{\textbf{4}} & \multicolumn{1}{c}{\textbf{1}} & \multicolumn{1}{c}{\textbf{2}} & \multicolumn{1}{c}{\textbf{3}} & \multicolumn{1}{c}{\textbf{4}} &  \\
GJ1001 & C & 689.8474 & 1067.8194 & 1083.7259 & 1545.6148 & 28 & 6 & -17 & -16 & 20 & 33 & 22 & 20 & 0.343 \\
GJ1002 & T & 293.9686 & 466.6685 & 469.5924 & 1049.9704 & 18 & -8 & 8 & -24 & 29 & 98 & 31 & 29 & 0.761 \\
LEP0011+5908 & T & 122.6580 & 476.6877 & 1102.8183 & 1856.7702 & 4 & -6 & -28 & 28 & 21 & 25 & 36 & 35 & 0.723 \\
GJ12 & T & 34.7181 & 280.9609 & 1014.9527 & 1828.8724 & 36 & 13 & -33 & -24 & 35 & 29 & 40 & 31 & 0.468 \\
GJ15B & T & 1.9447 & 301.9763 & 1009.9604 & 1851.8027 & -26 & -16 & 5 & 62 & 20 & 26 & 29 & 27 & 0.063 \\ \hline
\end{tabular}}
\tablecomments{In the instrument column, T denotes TRES and C denotes CHIRON. This table is published in its entirety in machine-readable format. A portion is shown here for guidance regarding its form and content. \newline}
\end{table*}

We note that our uncertainty estimates thus far do not account for the long-term instability of the spectrograph. Over the observation campaign, one zero-point offset occurred for CHIRON and two for TRES due to hardware and software changes to the instrument. To determine the amplitudes of these offsets, we create an ensemble of all our observations from each spectrograph, subtracting the error-weighted-mean radial velocity of each star to put the observations on a common zero point. We then use these ensembles to measure the amplitude of the radial-velocity discontinuities and apply a correction to subsequent measurements. For CHIRON, this offset is $+50$ ms$^{-1}$ at BJD $= 2458840$ days. For TRES, these offsets are $+45$ ms$^{-1}$ at BJD $= 2458700$ days and $-20$ ms$^{-1}$ at BJD $=2459218$ days.

To evaluate the stability of each spectrograph outside of these events, we consider the standard deviation of the most precise observations -- namely, observations with uncertainty estimates between 5 and 18 ms$^{-1}$. If our observations were unaffected by a noise floor, we would expect the standard deviation to fall somewhere within this 5--18 ms$^{-1}$ range. Note that the delta degrees of freedom (ddof) in this calculation is not 1; this would underestimate the standard deviation, as we have subtracted the error-weighted mean of each star. The ddof is also not equal to the number of stars; this would overestimate the standard deviation, as all observations are used to inform the error-weighted means, not just observations with small estimated uncertainties. We make a more reasonable estimate of the ddof by assuming that each star contributes $\sum w_{j} / \sum w_{i}$, where $w_j$ are the variance weights of the observations with estimated uncertainties below 18 ms$^{-1}$ and $w_i$ are the variance weights of all observations. The maximum value of this expression is 1, which occurs when the error-weighted-mean radial velocity is fully informed by observations with small estimated uncertainties. For CHIRON, our calculation includes 137 observations, has ddof = 47.4, and results in a standard deviation of 20 ms$^{-1}$. For TRES, we have 48 observations with small estimated uncertainties, \hbox{ddof = 16.2}, and a standard deviation of 21 ms$^{-1}$. As these results are greater than 18 ms$^{-1}$, it appears that instrumental instability is indeed setting a noise floor. These results are consistent with \cite{Winters2020}, who found typical RMS velocity residuals of 20 ms$^{-1}$ using orbital solutions for bright, slowly rotating low-mass M-dwarf binaries observed by TRES. As uncertainties below 20 ms$^{-1}$ may therefore be underestimated due to instrumental instability, we do not allow our estimates to be smaller than this floor.

Our radial-velocity measurements and uncertainties for all 200 stars are available in a machine-readable format, with the structure of this file shown in Table~\ref{tab:rvs}.

\needspace{6em}
\section{Data Analysis}
\label{sec:analysis}
\subsection{Identification and follow-up of candidate variables}
\label{sec:follow-up}
We use the metric $P(\chi^2)$ to evaluate whether each radial-velocity time series is consistent with an unvarying model given the uncertainties determined in the previous section. In this context, the chi-squared estimator is given by

\begin{equation}
    \chi^2 = \sum_{i=1}^{4} \frac{(v_{\mathrm{rad}, i} - \langle v_{\rm{rad}}\rangle)^2}{\delta_{{\rm RV},i}^2},
\end{equation}

\noindent with $N_{\rm obs}-1=3$ degrees of freedom. $\langle v_{\rm{rad}}\rangle$ denotes the error-weighted mean of the radial velocity time series. We flag a signal as significant if the variability is inconsistent with the null hypothesis with 99\% confidence ($P(\chi^2) < 1$\%). In other words, we are searching for statistically significant excess RV jitter; if such jitter is detected, further follow-up is necessary to verify that this jitter is due to a planet and not an astrophysical or statistical false positive. Given our sample size of 200 stars, we expect an average of two statistical false positives due to random chance.

We conducted a preliminary analysis of our data in early 2021, from which we identified two stars -- LHS 2899 and \hbox{G 125-34} -- as candidate variables based on this $P(\chi^2) < 1$\% cut. Based on the significance of the variability in the initial four observations, we proceeded to collect an additional 41 observations of LHS 2899 and 46 observations of \hbox{G 125-34} with TRES between 2021 April and 2022 May. At the time of the preliminary analysis, we had yet to obtain a fourth observation for some stars in the sample; our follow-up of LHS 2899 and G 125-34 therefore occurred concurrently with our main observation campaign, which was completed on 2022 May 11. When analyzing observations of all stars after data collection was completed, we found that the stability of the TRES zero point worsened in 2021--2022, motivating us to augment our 20 ms$^{-1}$ noise floor with a dynamic floor using the radial-velocity jitter in the TRES standard stars over each observing run (S. Quinn, private communication). These standards are quiet, Sun-like stars that are known to have very low intrinsic radial-velocity variation based on observations from more precise instruments. We use these standards to determine the RMS scatter in each observing run, which we add to our uncertainties in quadrature. In theory, we should also be able to correct for zero-point offsets using these standards; however, they are Sun-like stars and analyzed in bluer orders than our M-dwarf spectra, and we find that the offsets in the M dwarfs are not entirely commensurate with those seen in the standards.

With this refinement to our uncertainty estimation, the $P(\chi^2)$ in the initial four observations of LHS 2899 increased from $<1$\% to 3\%, no longer meeting our follow-up criterion. For the purposes of our statistical study, it is therefore not necessary to establish whether the variability of LHS 2899 is a planet or a false positive; however, for completeness we will nonetheless discuss it here. \hbox{G 125-34} retained its $P(\chi^2) < 1$\% designation and no other signals became significant with this change. $P(\chi^2)$ values for all stars are given in Table~\ref{tab:results}.

\begingroup
  \small
  \texttt{
  \begin{longtable*}{lllll|lllll}
    \caption{\textrm{Stellar properties and RV variability for our 200 stars}}\\
    \toprule
    Name & 2MASS ID & $M_*$ &
    $L_*$ & $P(\chi^2)$ & Name & 2MASS ID & $M_*$ &
    $L_*$ & $P(\chi^2)$ \\
    &  & [$M_\odot$] & [$L_\odot$] &  &  & & [$M_\odot$] & [$L_\odot$] & \\
    \endfirsthead
    \multicolumn{10}{c}{%
      \tablename\ \thetable\ -- \textit{Continued from previous page}%
    } \\
    \toprule
    Name & 2MASS ID & $M_*$ &
    $L_*$ & $P(\chi^2)$ & Name & 2MASS ID & $M_*$ &
    $L_*$ & $P(\chi^2)$ \\
    &  & [$M_\odot$] & [$L_\odot$] &  &  & & [$M_\odot$] & [$L_\odot$] & \\
    \endhead
    \multicolumn{10}{r}{\textit{Continued on next page}} \\
    \endfoot
    \endlastfoot
    GJ1001 & 00043643-4044020 & 0.262 & 0.0079 & 0.343 & LHS2385 & 11163766-2757186 & 0.186 & 0.0045 & 0.013 \\
GJ1002 & 00064325-0732147 & 0.115 & 0.0015 & 0.761 & LHS2395 & 11193058+4641437 & 0.112 & 0.0014 & 0.547 \\
LEP0011+5908 & 00113182+5908400 & 0.107 & 0.0012 & 0.723 & LHS2415 & 11285624+1010395 & 0.286 & 0.0091 & 0.929 \\
GJ12 & 00154919+1333218 & 0.249 & 0.0075 & 0.468 & LHS306 & 11310835-1457201 & 0.157 & 0.0035 & 0.078 \\
GJ15B & 00182549+4401376 & 0.157 & 0.0034 & 0.063 & SCR1138-7721 & 11381671-7721484 & 0.118 & 0.0017 & 0.291 \\
LHS112 & 00202922+3305081 & 0.119 & 0.0016 & 0.706 & SIP1141-3624 & 11412152-3624346 & 0.173 & 0.0041 & 0.322 \\
GJ1013 & 00313539-0552115 & 0.275 & 0.0088 & 0.250 & GJ442B & 11463269-4029476 & 0.164 & 0.0037 & 0.055 \\
L291-115 & 00331349-4733165 & 0.126 & 0.0020 & 0.667 & GJ445 & 11474143+7841283 & 0.252 & 0.0076 & 0.495 \\
GJ1014 & 00355557+1028352 & 0.132 & 0.0024 & 0.022 & GJ447 & 11474440+0048164 & 0.173 & 0.0041 & 0.593 \\
LHS1134 & 00432603-4117337 & 0.210 & 0.0057 & 0.175 & GJ1151 & 11505787+4822395 & 0.164 & 0.0038 & 0.122 \\
LHS1140 & 00445930-1516166 & 0.179 & 0.0044 & 0.619 & L758-107B & 12111697-1958213 & 0.272 & 0.0085 & 0.577 \\
GDR20049+6518 & 00492565+6518038 & 0.140 & 0.0028 & 0.609 & GJ465 & 12245243-1814303 & 0.274 & 0.0092 & 0.480 \\
GJ1025 & 01005643-0426561 & 0.198 & 0.0051 & 0.763 & GJ1158 & 12293453-5559371 & 0.233 & 0.0066 & 0.845 \\
GJ1028 & 01045368-1807292 & 0.136 & 0.0025 & 0.683 & LHS337 & 12384914-3822527 & 0.152 & 0.0033 & 0.712 \\
GJ1031 & 01081826-2848207 & 0.207 & 0.0055 & 0.478 & LHS2597 & 12393641-2658111 & 0.114 & 0.0015 & 0.621 \\
GJ1035 & 01195227+8409327 & 0.155 & 0.0034 & 0.686 & GJ480.1 & 12404633-4333595 & 0.185 & 0.0047 & 0.737 \\
GJ61B & 01365042+4123325 & 0.187 & 0.0047 & 0.166 & LHS2608 & 12421964-7138202 & 0.242 & 0.0071 & 0.395 \\
LP991-84 & 01392170-3936088 & 0.133 & 0.0023 & 0.449 & LHS2674a & 13065025+3050549 & 0.133 & 0.0023 & 0.384 \\
LHS5045 & 01525159-4805413 & 0.217 & 0.0061 & 0.611 & LHS2718 & 13200391-3524437 & 0.259 & 0.0079 & 0.151 \\
L173-19 & 02003830-5558047 & 0.275 & 0.0086 & 0.461 & LHS350 & 13225673+2428034 & 0.265 & 0.0080 & 0.442 \\
LHS1326 & 02021620+1020136 & 0.112 & 0.0013 & 0.092 & GJ1171 & 13303106+1909340 & 0.146 & 0.0031 & 0.728 \\
LHS1339 & 02054859-3010361 & 0.206 & 0.0057 & 0.154 & LHS2784 & 13424328+3317255 & 0.281 & 0.0088 & 0.203 \\
GJ105B & 02361535+0652191 & 0.263 & 0.0079 & 0.035 & LP911-56 & 13464607-3149258 & 0.104 & 0.0009 & 0.400 \\
GJ1050 & 02395066-3407557 & 0.287 & 0.0097 & 0.620 & GJ1179A & 13481341+2336486 & 0.122 & 0.0019 & 0.807 \\
SCR0246-7024 & 02460224-7024062 & 0.137 & 0.0026 & 0.675 & LTT5437 & 13571306-2922252 & 0.279 & 0.0091 & 0.101 \\
LHS1443 & 02463486+1625115 & 0.101 & 0.0009 & 0.724 & SSS1358-3938 & 13580529-3937545 & 0.129 & 0.0023 & 0.713 \\
LP831-1 & 02543950-2215584 & 0.246 & 0.0074 & 0.540 & LHS2830 & 13581392+1234438 & 0.220 & 0.0061 & 0.780 \\
LHS1481 & 02581021-1253066 & 0.184 & 0.0048 & 0.731 & G165-58 & 14155637+3616368 & 0.241 & 0.0071 & 0.351 \\
LTT1445A & 03015142-1635356 & 0.258 & 0.0080 & 0.956 & GJ545 & 14200739-0937127 & 0.263 & 0.0079 & 0.036 \\
LHS1490 & 03020638-3950516 & 0.111 & 0.0014 & 0.993 & LHS2899 & 14211512-0107199 & 0.237 & 0.0067 & 0.030 \\
GJ1055 & 03090015+1001257 & 0.133 & 0.0024 & 0.095 & LEP1422-7023 & 14221943-7023371 & 0.121 & 0.0019 & 0.541 \\
GJ1053 & 03105861+7346189 & 0.128 & 0.0023 & 0.028 & LEP1455+3006 & 14551146+3006454 & 0.170 & 0.0040 & 0.914 \\
GJ1057 & 03132299+0446293 & 0.160 & 0.0035 & 0.632 & GJ2112 & 15221293-2749436 & 0.180 & 0.0043 & 0.356 \\
LHS1516 & 03141241+2840411 & 0.108 & 0.0012 & 0.640 & GJ585 & 15235112+1727569 & 0.171 & 0.0041 & 0.560 \\
GJ1059 & 03230175+4200269 & 0.124 & 0.0021 & 0.419 & GJ589B & 15352039+1743045 & 0.129 & 0.0023 & 0.586 \\
LHS176 & 03353849-0829223 & 0.119 & 0.0017 & 0.725 & GJ589A & 15352059+1742470 & 0.296 & 0.0103 & 0.961 \\
GJ1061 & 03355969-4430453 & 0.123 & 0.0019 & 0.946 & GJ590 & 15363450-3754223 & 0.211 & 0.0056 & 0.460 \\
L228-92 & 03385590-5234107 & 0.146 & 0.0031 & 0.257 & GJ1194A & 15400352+4329396 & 0.295 & 0.0103 & 0.486 \\
LHS1593 & 03472091+0841464 & 0.134 & 0.0025 & 0.283 & GJ1194B & 15400374+4329355 & 0.204 & 0.0050 & 0.962 \\
GJ1065 & 03504432-0605400 & 0.198 & 0.0051 & 0.614 & GJ609 & 16025098+2035218 & 0.250 & 0.0074 & 0.727 \\
LP357-56 & 03544620+2416246 & 0.116 & 0.0013 & 0.746 & GJ611B & 16045093+3909359 & 0.145 & 0.0029 & 0.115 \\
2MA0406-0534 & 04060688-0534444 & 0.224 & 0.0061 & 0.164 & GJ618B & 16200321-3731485 & 0.164 & 0.0034 & 0.079 \\
LHS1629 & 04063732+7916012 & 0.134 & 0.0024 & 0.242 & LHS3241 & 16463154+3434554 & 0.108 & 0.0011 & 0.928 \\
GJ1068 & 04102815-5336078 & 0.129 & 0.0023 & 0.597 & LP154-205 & 16475517-6509116 & 0.168 & 0.0039 & 0.355 \\
GJ1072 & 04505083+2207224 & 0.139 & 0.0026 & 0.499 & GJ643D & 16552527-0819207 & 0.210 & 0.0056 & 0.226 \\
GJ1073 & 04523448+4042255 & 0.212 & 0.0057 & 0.935 & LHS3262 & 17032384+5124219 & 0.180 & 0.0044 & 0.641 \\
LHS1723 & 05015746-0656459 & 0.166 & 0.0039 & 0.931 & GJ1214A & 17151894+0457496 & 0.178 & 0.0042 & 0.828 \\
LHS1731 & 05032009-1722245 & 0.290 & 0.0097 & 0.798 & GJ1220 & 17311725+8205198 & 0.158 & 0.0035 & 0.494 \\
UPM0505+4414 & 05050591+4414037 & 0.145 & 0.0030 & 0.630 & LHS3324 & 17460465+2439049 & 0.278 & 0.0092 & 0.919 \\
G86-28 & 05103956+2946479 & 0.261 & 0.0082 & 0.858 & GJ693 & 17463427-5719081 & 0.280 & 0.0092 & 0.413 \\
LEHPM2-1009 & 05273058-5129158 & 0.129 & 0.0022 & 0.195 & BARNARDS & 17574849+0441405 & 0.155 & 0.0033 & 0.168 \\
GJ203 & 05280015+0938382 & 0.231 & 0.0065 & 0.089 & GJ1223 & 18024624+3731048 & 0.139 & 0.0027 & 0.579 \\
LHS5108 & 05325194+3349474 & 0.202 & 0.0053 & 0.471 & LP449-10 & 18064856+1720472 & 0.196 & 0.0051 & 0.951 \\
GJ213 & 05420897+1229252 & 0.221 & 0.0061 & 0.243 & G140-51 & 18163154+0452456 & 0.174 & 0.0040 & 0.066 \\
GJ2045 & 05421271-0527567 & 0.122 & 0.0019 & 0.712 & LHS461B & 18180345+3846359 & 0.165 & 0.0038 & 0.904 \\
LEP0556+1144 & 05565722+1144333 & 0.125 & 0.0020 & 0.597 & GJ712 & 18220671+0620376 & 0.294 & 0.0100 & 0.244 \\
LHS1805 & 06011106+5935508 & 0.275 & 0.0085 & 0.470 & GJ1227 & 18222719+6203025 & 0.162 & 0.0037 & 0.022 \\
LHS1809 & 06022918+4951561 & 0.134 & 0.0024 & 0.693 & SCR1841-4347 & 18410977-4347327 & 0.111 & 0.0013 & 0.398 \\
GJ1088 & 06105288-4324178 & 0.296 & 0.0097 & 0.088 & GJ1230B & 18410981+2447195 & 0.195 & 0.0042 & 0.069 \\
G192-22 & 06140240+5140081 & 0.264 & 0.0083 & 0.667 & LP867-15 & 18421107-2328582 & 0.259 & 0.0081 & 0.262 \\
LP779-34 & 06151198-1626152 & 0.190 & 0.0049 & 0.067 & GJ725B & 18424688+5937374 & 0.265 & 0.0084 & 0.470 \\
L308-57 & 06210665-4905379 & 0.172 & 0.0040 & 0.301 & LHS5341 & 18430697-5436481 & 0.240 & 0.0071 & 0.033 \\
GJ232 & 06244132+2325585 & 0.152 & 0.0032 & 0.482 & G184-31 & 18495449+1840295 & 0.162 & 0.0037 & 0.497 \\
SCR0642-6707 & 06422703-6707193 & 0.110 & 0.0013 & 0.731 & GJ732A & 18533991-3836442 & 0.289 & 0.0093 & 0.368 \\
GJ1092 & 06490542+3706533 & 0.177 & 0.0042 & 0.027 & GJ1232 & 19095098+1740074 & 0.189 & 0.0049 & 0.077 \\
GJ1093 & 06592868+1920577 & 0.123 & 0.0018 & 0.822 & LEP1916+8413 & 19162483+8413411 & 0.150 & 0.0031 & 0.256 \\
G107-48 & 07073776+4841138 & 0.163 & 0.0037 & 0.720 & GJ754.1B & 19203346-0739435 & 0.259 & 0.0087 & 0.187 \\
L136-37 & 07205204-6210118 & 0.265 & 0.0083 & 0.946 & GJ754 & 19204795-4533283 & 0.175 & 0.0042 & 0.022 \\
SCR0736-3024 & 07365666-3024160 & 0.196 & 0.0051 & 0.535 & LHS475 & 19205439-8233170 & 0.275 & 0.0085 & 0.107 \\
GJ283B & 07401922-1724449 & 0.101 & 0.0008 & 0.737 & GJ1235 & 19213867+2052028 & 0.194 & 0.0050 & 0.084 \\
LHS1950 & 07515138+0532572 & 0.154 & 0.0034 & 0.255 & GJ1236 & 19220206+0702310 & 0.223 & 0.0064 & 0.066 \\
GJ1103 & 07515465-0000117 & 0.192 & 0.0049 & 0.176 & GJ1238 & 19241634+7533121 & 0.121 & 0.0018 & 0.073 \\
GJ1105 & 07581269+4118134 & 0.284 & 0.0090 & 0.653 & G125-34 & 19484080+3555178 & 0.237 & 0.0067 & 0.008 \\
GJ299 & 08115757+0846220 & 0.138 & 0.0027 & 0.428 & GJ770C & 19542064-2356398 & 0.166 & 0.0036 & 0.719 \\
GJ300 & 08124088-2133056 & 0.282 & 0.0091 & 0.117 & GJ1248 & 20035098+0559440 & 0.267 & 0.0090 & 0.332 \\
LHS2025 & 08313011+7303459 & 0.279 & 0.0088 & 0.620 & GJ774B & 20040195-6535586 & 0.220 & 0.0062 & 0.858 \\
GJ2070 & 08342587-0108391 & 0.246 & 0.0074 & 0.103 & GJ1253 & 20260528+5834224 & 0.158 & 0.0034 & 0.321 \\
LEP0840+3127 & 08401597+3127068 & 0.299 & 0.0098 & 0.130 & GJ1251 & 20280382-7640164 & 0.169 & 0.0039 & 0.979 \\
LEP0844-4805 & 08443891-4805218 & 0.200 & 0.0053 & 0.148 & GJ1256 & 20403364+1529572 & 0.189 & 0.0048 & 0.096 \\
GJ324B & 08524084+2818589 & 0.278 & 0.0089 & 0.957 & LP816-60 & 20523304-1658289 & 0.238 & 0.0069 & 0.743 \\
LHS2088 & 08595604+7257364 & 0.158 & 0.0035 & 0.573 & LHS3593 & 20533304+1037020 & 0.180 & 0.0044 & 0.014 \\
LP60-179 & 09025284+6803464 & 0.265 & 0.0081 & 0.038 & GJ810B & 20553706-1403545 & 0.140 & 0.0028 & 0.078 \\
GJ1123 & 09170532-7749233 & 0.224 & 0.0063 & 0.686 & GJ2151 & 21031390-5657479 & 0.242 & 0.0071 & 0.751 \\
LEP0921-0219 & 09214812-0219433 & 0.272 & 0.0090 & 0.679 & LEP2124+4003 & 21243234+4003599 & 0.130 & 0.0022 & 0.660 \\
GJ359 & 09410199+2201291 & 0.145 & 0.0030 & 0.782 & LHS510 & 21304763-4042290 & 0.205 & 0.0053 & 0.445 \\
GJ1128 & 09424635-6853060 & 0.176 & 0.0043 & 0.258 & WT795 & 21362532-4401005 & 0.197 & 0.0052 & 0.162 \\
LHS5156 & 09424960-6337560 & 0.210 & 0.0057 & 0.788 & LHS512 & 21384369-3339555 & 0.280 & 0.0088 & 0.075 \\
LHS272 & 09434633-1747066 & 0.154 & 0.0036 & 0.478 & LEP2146+3813 & 21462206+3813047 & 0.178 & 0.0043 & 0.093 \\
GJ1129 & 09444731-1812489 & 0.299 & 0.0099 & 0.318 & LP698-42 & 21471744-0444406 & 0.157 & 0.0035 & 0.539 \\
LHS2224 & 10092996+5117197 & 0.181 & 0.0044 & 0.831 & LHS516 & 21565513-0154100 & 0.144 & 0.0030 & 0.060 \\
GJ1132 & 10145184-4709244 & 0.192 & 0.0049 & 0.765 & LHS3746 & 22022935-3704512 & 0.251 & 0.0074 & 0.275 \\
LEP1015+1729 & 10155390+1729271 & 0.287 & 0.0091 & 0.367 & GJ1265 & 22134277-1741081 & 0.168 & 0.0040 & 0.699 \\
LEHPM2-2758 & 10384782-8632441 & 0.252 & 0.0075 & 0.097 & GJ1270 & 22294885+4128479 & 0.259 & 0.0078 & 0.381 \\
GJ1134 & 10413809+3736397 & 0.205 & 0.0054 & 0.408 & L645-74B & 22382544-2921244 & 0.254 & 0.0076 & 0.066 \\
LHS288 & 10442131-6112384 & 0.106 & 0.0012 & 0.959 & LHS3844 & 22415815-6910089 & 0.151 & 0.0031 & 0.496 \\
LHS2303 & 10442927-1838063 & 0.125 & 0.0021 & 0.853 & GJ1277 & 22562466-6003490 & 0.172 & 0.0040 & 0.573 \\
LHS2310 & 10473868-7927458 & 0.251 & 0.0075 & 0.605 & LEHPM2-2163 & 23303802-8455189 & 0.138 & 0.0027 & 0.240 \\
GJ402 & 10505201+0648292 & 0.282 & 0.0091 & 0.056 & GJ1286 & 23351050-0223214 & 0.118 & 0.0017 & 0.160 \\
GJ403 & 10520440+1359509 & 0.278 & 0.0087 & 0.164 & LHS547 & 23365227-3628518 & 0.171 & 0.0040 & 0.968 \\
LHS296 & 11011965+0300171 & 0.164 & 0.0037 & 0.692 & GJ905 & 23415498+4410407 & 0.140 & 0.0026 & 0.125 \\
\label{tab:results}
  \end{longtable*}}
\endgroup

\begin{figure*}[h]
    \centering
    \vspace{-0.9cm}
    \makebox[\textwidth][c]{\includegraphics[width=0.85\textwidth]{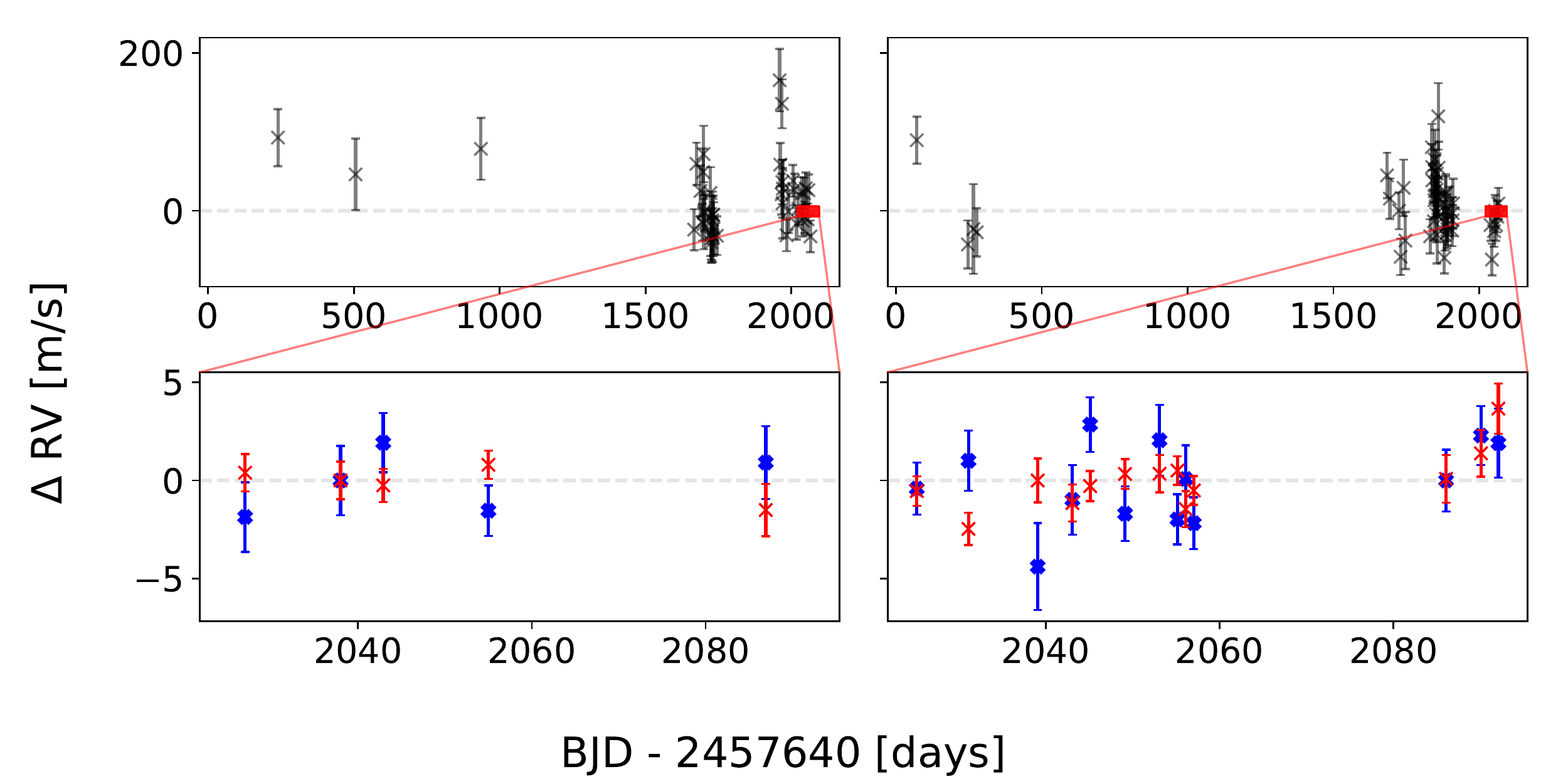}}
    \caption{Radial-velocity time series for the two candidate RV variables. The left panels show LHS 2899 and the right panels show \hbox{G 125-34}, while the upper panels show TRES RVs and the lower panels show MAROON-X RVs. The lower panels cover a much smaller range in time and radial velocity; the overlap is indicated by the red rectangles. Radial velocities from the blue arm of MAROON-X are shown in blue with thick crosses and from the red arm in red with thin crosses. The red arm achieves uncertainties as low as 70 cms$^{-1}$. The lack of radial-velocity variability in the extreme-precision MAROON-X observations refutes all possible orbital solutions consistent with the variability in the TRES data.}
    \label{fig:cand}
    \centering
    \makebox[\textwidth][c]{\includegraphics[width=1.15\textwidth]{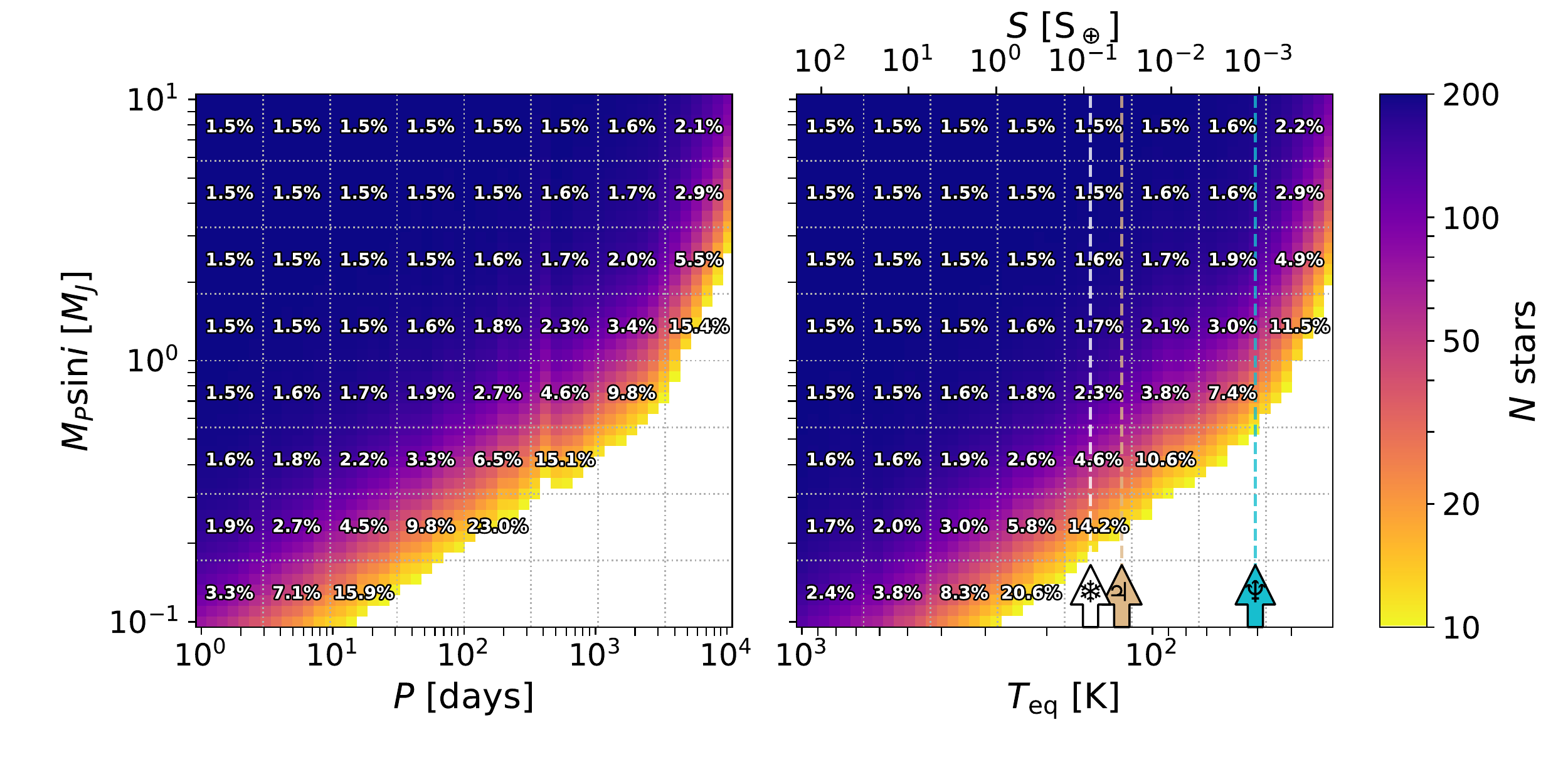}}
    \caption{The completeness of our survey, annotated with 95\%-confidence upper limits on the occurrence rate of giant planets around low-mass M dwarfs as a function of planetary mass and period/instellation/zero-albedo equilibrium temperature under the assumption of circular orbits. In regions where we are highly sensitive (i.e., short periods and large masses), we constrain the occurrence rate to $<1.5\%$ with 95\% confidence. The colorbar indicates $N_{\rm eff}$, the effective number of stars around which we are sensitive to the hypothetical planet. The arrows in the righthand plot show instellations equivalent to that of the water snow line \citep{Podolak2004} (white), Jupiter (brown), and Neptune (blue). We are sensitive to $M_{\rm P}$sin$i = 1$M$_{\rm J}$ planets at the water snow line for 83\% of stars (occurrence rate $< 1.8$\% with 95\% confidence) and at Jupiter-like instellations for 70\% of stars (occurrence rate $<$ 2.1\%). We remain sensitive to $M_{\rm P}$sin$i = 1$M$_{\rm J}$ planets at Neptune-like instellations around 19\% of stars (occurrence rate $<$ 7.7\%). \vspace{-2cm}}
    \label{fig:occ}
\end{figure*}

For both \hbox{G 125-34} and LHS 2899, the significance of the candidate variability increased with the TRES follow-up observations, with a final $P(\chi^2)$ = 4.3$\times10^{-5}$ for \hbox{G 125-34} and $P(\chi^2)$ = 1.0$\times10^{-5}$ for LHS 2899. A periodogram analysis yielded candidate periods for orbital solutions, although the precision and quantity of the data were insufficient to definitively prove or disprove the presence of a planet. The preferred planetary solutions were a 0.22M$_{\rm J}$ planet on a 77-day orbit for \hbox{G 125-34} and a 0.70M$_{\rm J}$ planet on a 337-day orbit for LHS 2899.

To reach definitive conclusions about the dispositions of these candidates, we obtained 5 observations of LHS 2899 and 13 observations of \hbox{G 125-34} using the MAROON-X spectrograph \citep{Seifahrt2018}, an extreme-precision radial-velocity instrument on the \hbox{8.1 m} Gemini-North telescope, over a 2-month interval. These observations were reduced by the MAROON-X team using the SERVAL pipeline \citep{Zechmeister2018} and are shown in Figure~\ref{fig:cand}. The final observation of LHS 2899 and the final three observations of \hbox{G 125-34} were taken in a different MAROON-X observing run from the prior observations, with a zero-point offset of $-1.5 \pm 1.0 $ms$^{-1}$ for the blue arm and $+1.5 \pm 1.0 $ms$^{-1}$ for the red arm; the uncertainty in this offset is responsible for the larger error bars for these later observations. The radial-velocity time series for both stars are flat at the ms$^{-1}$ level, definitively ruling out all orbital solutions consistent with the TRES data. LHS 2899 and \hbox{G 125-34} are statistical false positives, with their significance in the TRES follow-up observations likely the result of non-Gaussian outliers. There are no giant planet detections in our 200-star sample.

\subsection{Occurrence rate constraints}  \label{sec:occurrence}

For a given star, $j$, the probability that we do not detect a planet is $P_{\rm null}(\mu)_j = 1-\mu\times C_j$, where $\mu$ is the planetary occurrence rate in a given bin and $C$ is the survey completeness in that bin. The probability that we do not detect any planets around any of our stars is then the product

\begin{equation}
    P_{\rm null}(\mu) = \prod_{j} (1-\mu C_j).
\end{equation}

\noindent To determine a 95\%-confidence limit on the occurrence rate, we numerically solve for $\mu$ given that $P_{\rm null}(\mu) = 0.05$. While we use this form of the equation to generate our Figure~\ref{fig:occ}, note that in the limit where the binomial approximation \hbox{$(1-\mu)^{C_j} \approx 1-\mu C_j$} applies (i.e., where $\mu < 1$ and $\mu C_j \ll 1$, which is appropriate for this study), this equation can be rewritten as \hbox{$P_{\rm null}(\mu)=(1-\mu)^{N_{\rm eff}}$,} where $N_{\rm eff}$, the effective sample size, is given by $N_{\rm eff} = \sum_j C_j$. This form is a convenient simplification, as it only requires the average completeness instead of the completeness for each star in the sample, and it motivates our use of $N_{\rm eff}$ as the color bar axis in this figure.

\begin{figure*}[t]
    \centering
    \makebox[\textwidth][c]{\includegraphics[width=0.98\textwidth]{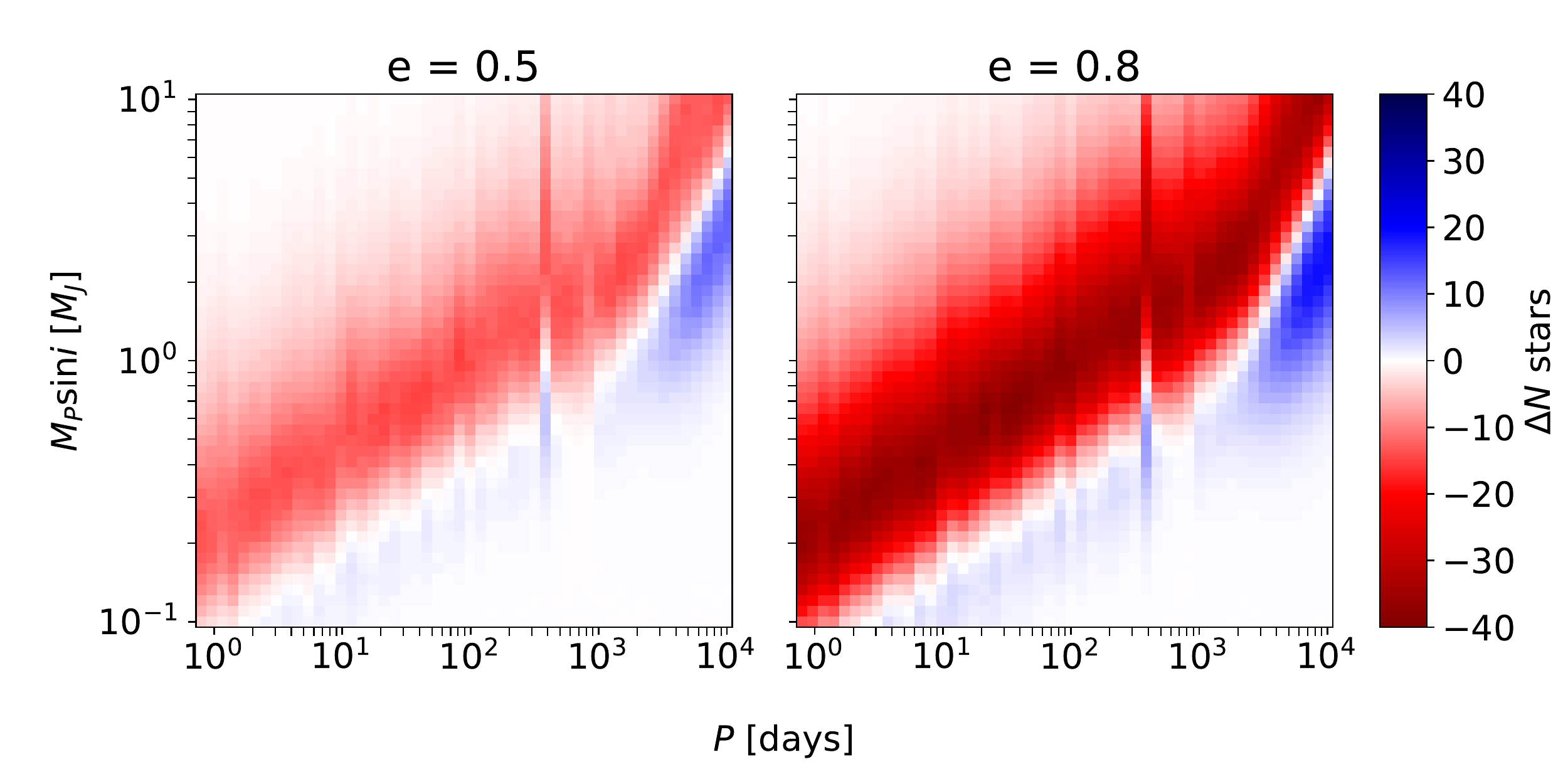}}
    \vspace{-0.7cm} 
    \caption{The difference in survey sensitivity between the assumption of circular orbits and the assumption that all planets have $e=0.5$ (left panel) or $e=0.8$ (right panel). This difference is given in terms of $\Delta N_{\rm eff}$, the change in the number of stars around which we could detect the planet, with a positive value indicating an increase in the effective number of stars for the eccentric case over the circular case. Given that our total sample contains 200 stars, the circular orbits assumption is generally appropriate for most regions of parameter space; the largest percent decrease in effective sample size is only 13\% and 30\% for the $e=0.5$ and $e=0.8$ assumptions, respectively. There are also previously inaccessible regions of parameter space to which we gain sensitivity when considering eccentric planets (in particular, the regime of super-Jupiters on wide orbits).}
    \label{fig:ecc}
\end{figure*}

We conduct an injection and recovery analysis to determine $C$ as a function of $M_{\rm P}$sin$i$ and orbital period. We derive our occurrence rate constraints under the assumption of circular orbits, following the precedent of past works \citep[e.g.,][]{Howard2010a, Mayor2011}. This assumption has been shown to be reasonable for $e < 0.5$ \citep{Endl_2002, Cumming2010}. While many eccentric cold Jupiters meet this criterion \citep{Buchhave2018}, more highly eccentric giant planets are known \citep[e.g.,][]{Robertson2012, Moutou2015}. Notably, giant planet formation via gravitational instability may produce large orbital eccentricities \hbox{\citep{Jennings2021}}. We consider the limitations of the circular orbits assumption in the following section.

We consider 100 periods spaced logarithmically between $10^{-1}$ and $10^5$ days and 100 values of $M_{\rm P}$sin$i$ spaced logarithmically between $10^{-1}$ and $10^{1}$ M$_{\rm J}$. For each mass and period, we generate 100 artificial radial-velocity time series for each star using the observation times and radial-velocity uncertainties of the real data, drawing phases from a random uniform distribution. For each hypothetical planet, every star is assigned a completeness $C$ between 0 and 1, indicating the fraction of trials in which the simulated observations were variable at the $P(\chi^2) < 1\%$ level. The sample size $N_{\rm eff}$ is the sum of these fractions. This completeness assumes that we would gather sufficient MAROON-X follow-up observations to verify that any signal flagged at $P(\chi^2) < 1\%$ significance is a true planet.

Our injection and recovery results are shown in Figure~\ref{fig:occ}, which also includes the period axis recast in terms of zero-albedo equilibrium temperature, $T_{\rm eq} = [L/(16\pi\sigma a^2)]^{1/4}$, and instellation, $S = 4\sigma T_{\rm eq}^4$. This transformation requires the masses and luminosities of each star; we adopt the masses from \cite{Winters2021}, which are based on the $K$-band relation from \cite{Benedict2016}, and estimate the bolometric luminosities from the photometry collated in \cite{Winters2021} and distances from Gaia parallaxes \citep{Gaia2018, Gaia2021, Lindegren2021}. In particular, we consider the bolometric corrections from both \cite{Pecaut2013} and \hbox{\cite{Mann2015}}, adopting the average of the luminosities calculated from these two methods. These masses and luminosities are included in Table~\ref{tab:results}.

There are two regimes in Figure~\ref{fig:occ}, with a transition around $T_{\rm eq} \approx 50$ K. At short periods, the sensitivity is determined by the radial-velocity semiamplitude of the orbit. The slope of the transition region between detectable and undetectable planets represents the interplay between the radial-velocity semiamplitude and typical radial-velocity uncertainties. At long periods, the sensitivity drops off rapidly due to the finite length of the observing campaign.

\subsection{Eccentric orbits}
\label{sec:ecc}
To quantify the impact of the circular orbit assumption, we repeat the analysis of the previous section but under the assumption that all planets have $e=0.5$ in one trial or $e=0.8$ in another. For each simulated planet, we draw the argument of periastron from a random uniform distribution. Figure~\ref{fig:ecc} shows the change in sensitivities compared to the circular orbits case in terms of $\Delta N_{\rm eff}$, the change in the number of stars around which the planet is detectable. We find that $\Delta N_{\rm eff}$ is generally small; the sample size is never decreased by more than 15 stars in the $e = 0.5$ trial and 39 stars in the $e = 0.8$ trial. There are other regions of parameter space in which the sample size is increased by similar amounts. Moreover, the percent decrease in effective sample size does not exceed \hbox{13\%} or 30\%, respectively. Even in the limit where all giant planets are highly eccentric, the circular orbits assumption will produce reasonable occurrence rate constraints.

Compared to a planet on a circular orbit, an eccentric planet achieves higher radial-velocity semiamplitudes but spends more time at $\Delta$RVs near zero. As we approach the sensitivity limit from the detectable side, the decreased likelihood of observing the eccentric orbit in a peak/trough renders a previously detectable planet undetectable. As we approach the sensitivity limit from the undetectable side, the increased amplitude of the peaks/troughs renders a previously undetectable planet detectable.

\subsection{Summary of analysis}
To summarize the above reduction and analysis, we do not detect any giant planets in our sample of 200 low-mass, inactive M dwarfs. More specifically, our four measurements for each star do not vary in excess of our $P(\chi^2)$ = 1\% detection limit for 198 stars, and we initially flagged the remaining two stars as candidate RV variables. We collected additional observations of these two candidates with TRES, but were unable to conclusively establish the provenance of the signals. We then obtained observations of the two stars using MAROON-X. The MAROON-X radial velocities show no variation at the ms$^{-1}$ level over a 2-month interval, indicating that the candidate variability from the TRES observations are statistical false positives. After refuting these candidates, none of our stars exhibit radial-velocity variability that exceeds our detection threshold. Nevertheless, our injection and recovery analysis indicates that we are very sensitive to a variety of giant planets (Figure~\ref{fig:occ}). We can therefore place strong upper limits on the occurrence rate of these planets, as we discuss in the following section. The median mass of stars in our sample is 0.18M$_\odot$, with a median radial-velocity precision of 29 ms$^{-1}$ and a median observation baseline of 3.1 years.

\section{Discussion}
\label{sec:discussion}
\subsection{Occurrence of warm Jupiters}
\label{sec:warm}
As shown in Figure~\ref{fig:occ}, we have nearly complete sensitivity to $M_{\rm P}$sin$i> 1$M$_{\rm J}$ planets out to the water snow line. In regions of complete sensitivity, we place a 68\%-confidence upper limit of 0.57\% and a 95\%-confidence upper limit of 1.5\%. Jupiter- and super-Jupiter-mass planets interior to the snow line of low-mass M dwarfs are therefore exceedingly rare.

How does this compare to studies of all types of M dwarfs? The \citet{Bonfils_2013} sample contains one planet in this regime: Gliese 876 b \citep{Marcy1998, Delfosse1998}. While that work published their survey sensitivity in specific bins that differ from the one currently under discussion (their Table 11), inspection of their Figure 15 indicates that they are nearly complete across this parameter range, with $N_{\rm eff}$ equal to roughly 96 stars. If the true occurrence rate of such planets around all M dwarfs was equal to our 68\%-confidence upper limit of 0.57\%, binomial probability yields a 32\% chance that \citet{Bonfils_2013} would detect one planet in this bin, with a 58\% chance of detecting zero and a 10\% chance of detecting two or more planets. Our results are therefore in reasonable agreement with \citet{Bonfils_2013} without invoking a decrease in warm Jupiter occurrence around low-mass M dwarfs relative to all M dwarfs, although our occurrence constraints are tighter given that our sample is twice as large. We also note that Gliese 876 has a mass of 0.37M$_\odot$; i.e., while it is more massive than the M dwarfs in our sample, it is not a particularly massive M dwarf. That said, a decrease in warm Jupiter occurrence around low-mass M dwarfs relative to all M dwarfs is also consistent with our data.

Other constraints on the occurrence rate of giant planets around early M dwarfs are available from transit surveys, although they are limited to hot Jupiters and their bins are not as directly comparable to ours due to the observable of these studies being radius, not mass. \citet{Gan2023} found an occurrence rate of 0.27$\pm$0.09\% for hot Jupiters around early ($0.45 \leq M_* \leq 0.65$) M dwarfs from TESS, with this statistic defined over the range 7$R_\oplus \leq R_{\rm P} \leq 2 R_{\rm J}, 0.8 \leq P \leq 10$ days. A similar occurrence rate of such planets around our low-mass M dwarfs is fully consistent with our null detection.

\subsection{Occurrence of giant planets at the snow line}
Planet formation theories predict an enhancement in giant planet occurrence at the water snow line \citep{Pollack1996, Ida2008, Morbidelli2015}, a phenomenon that has been observed around Sun-like stars; specifically, \citet{Fulton2021} found that giant planet occurrence in the California Legacy Survey of FGKM stars follows a broken power law, with a peak that approximately coincides with the location of the snow line and which represents a factor of 4 increase in occurrence relative to giant planets interior to the snow line. While this sample does contain a small number of M dwarfs, it is dominated by FGK stars, with 83\% of stars in the sample having stellar masses above 0.6M$_\odot$ and 98\% having stellar masses above 0.3M$_\odot$ (see Figure 4 of \citealt{Rosenthal2021}).

As an aside, we note that \citet{Fulton2021} do not make specific statements about the occurrence of giant planets at the snow line of their M dwarfs. In their Figure 7, they do present occurrence rates as a function of stellar mass for giant plants that are more massive than Saturn and that orbit between 1--5 au, including a bin for 0.3--0.5M$_\odot$ M dwarfs. In this bin, they find an occurrence rate of roughly 5\%, with the 68\%-confidence interval ranging from 5\% to 12\%. However, this 1--5 au bin represents much lower instellations for M dwarfs than for Sun-like stars. Using the scaling relation $a_{\rm snow} = 2.7 (M_*/$M$_\odot)^{1.14}$ au from \citet[][adapted from \citealt{Mulders2015}]{Childs2022}, the snow line of a 0.4M$_\odot$ star occurs at 0.95 au and decreases further to 0.68 au for a 0.3M$_\odot$ star. Therefore, while the \hbox{1--5 au} bin is centered on the snow line for Sun-like stars, it does not represent the occurrence rate of giant planets at the snow line of M dwarfs. Rather, it represents the occurrence rate of giant planets on wider orbits.

To investigate the occurrence rate of giant planets at the snow line of low-mass M dwarfs in our sample, we consider a bin corresponding to zero-albedo equilibrium temperatures of $100$ K $< T_{\rm eq} < 150$ K, which we subdivide into mass categories of sub-Jupiters (0.3M$_{\rm J} < M_{\rm P}$sin$i < 0.8$M$_{\rm J}$), Jupiters (0.8M$_{\rm J} < M_{\rm P}$sin$i < 3$M$_{\rm J}$), and super-Jupiters (3M$_{\rm J} < M_{\rm P}$sin$i < 10$M$_{\rm J}$). Assuming uniform occurrence in log($T_{\rm eq}$) and log($M_{\rm P}$sin$i$) across each bin, our survey yields 95\%-confidence upper limits of $<$1.5\% for super-Jupiters, $<$1.7\% for Jupiters, and $<$4.4\% for sub-Jupiters in this region. If we combine the sub-Jupiters and Jupiters into a single 0.3M$_{\rm J} < M_{\rm P}$sin$i < 3$M$_{\rm J}$ mass bin, we obtain a 95\%-confidence constraint of $<$2.4\%. While this latter bin averages over a strong gradient in our survey completeness, \cite{Fulton2021} showed that the assumption of uniform occurrence in log($M_{\rm P}$sin$i$) holds between 0.1M$_{\rm J}$ and 3M$_{\rm J}$ in their survey of Sun-like stars. This assumption would therefore be well justified if the formation of all giant planets is less efficient around low-mass M dwarfs, but poorly justified if Jupiter analogs are inhibited around low-mass M dwarfs but sub-Jupiter formation remains efficient (which we discuss in Section~\ref{sec:dist}).

Our survey indicates that Jupiter analogs -- planets with instellations and masses comparable to Jupiter in our solar system -- are rare around low-mass M dwarfs. Moreover, if the \citet{Fulton2021} Sun-like star result holds for low-mass M dwarfs (that is, if giant planets are 4 times more common near the snow line than on interior orbits), our null detection of giant planets in this region implies an even stronger constraint on warm Jupiters than that presented in the previous section. For our Jupiter bin, we find a 68\%-confidence upper limit of 0.66\%. A fourth of this value is 0.17\%; of course, it is also possible that the Sun-like occurrence rate distribution does not hold for low-mass M dwarfs, as we discuss in Section~\ref{sec:dist}.

How do our findings at and beyond the snow line compare with other works? It is notable that while few giant planets have been detected around low-mass M dwarfs using radial velocities, others have been inferred through microlensing events \citep[e.g.,][]{Skowron2015, Novati2018, Ryu2019}. However, comparing microlensing and radial-velocity statistics is challenging due to the different observables available to each technique. \citet{Clanton2014a} attempted to circumvent these issues by developing a mapping between microlensing and radial-velocity parameters that marginalizes over the unknown physical parameters in the microlensing sample. Using these methods, \citet{Clanton2014} synthesized radial-velocity and microlensing surveys to find an occurrence rate of $2.9^{+1.3}_{-1.5}$\% for M-dwarf planets with $P = 10^0 - 10^4$ days and 1M$_{\rm J} < M_{\rm P}$sin$i < 13$M$_{\rm J}$. For periods less than $10^2$ days, their statistics are fully informed by the radial-velocity results of \citet{Bonfils_2013}, with no contribution from the microlensing sample. As discussed in Section~\ref{sec:warm}, \citet{Bonfils_2013} yield an occurrence rate of 1\% (that is, 1/96) for these warm Jupiters; subtracting this from the 2.9\% found by \citet{Clanton2014} implies a roughly 1.9\% occurrence rate for $P = 10^2 - 10^4$ days and 1M$_{\rm J} < M_{\rm P}$sin$i < 13$M$_{\rm J}$. Assuming uniform occurrence in log($P$) and log($M_{\rm P}$sin$i$) across the bin, our survey obtains $N_{\rm eff}=148$ for giant planets in this regime, producing a 68\%-confidence upper limit of 0.77\% and a 95\%-confidence upper limit of 2.0\% (although note that our sensitivity drops precipitously towards the long-period edge of this bin). Given the large uncertainty in the \citet{Clanton2014} result, the lack of $M_{\rm P}$sin$i > 1$M$_{\rm J}$ planet detections in our study is not necessarily in tension with their occurrence rate. However, those authors note that their value is actually more like a lower limit on occurrence rate, as they use lower limits in the $P = 10^2 - 10^3$-day range where microlensing sensitivity declines, and they claim their findings are consistent with the nearly twice as large measurement of this occurrence rate from \citet{Montet2014}, another radial-velocity survey. As these studies focus on more massive M dwarfs (recall that the microlensing lens distribution is centered at $M_*=0.5$M$_\odot$, while the mean stellar mass in the \citealt{Bonfils_2013} sample is 0.35M$_\odot$), our lack of detections of any giant planets beyond the snow line may be indicative of a decrease in cold Jupiter occurrence around low-mass M dwarfs relative to all M dwarfs.

It is difficult to compare our results directly with \citet{Montet2014}, as they report their occurrence rate only in the bin $1M_{\rm J} < M_{\rm P} < 13M_{\rm J}$ out to 20 au. For a 0.2M$_\odot$ star, 20 au corresponds to a period of 7$\times$10$^{4}$ days; our survey is insensitive to planets at such long periods. We have similar difficulty comparing with the \citet{Fulton2021} occurrence rate of 5\% for 100--6000M$_\oplus$ planets that orbit 0.3--0.5M$_\odot$ M dwarfs between 1--5 au; we have no sensitivity to a 100M$_\oplus$ planet at 5 au. While we could compare with these surveys by assuming a specific function for the distribution of planets, as done in \citet{Montet2014}, we argue in Section~\ref{sec:dist} that the population of planets around low-mass M dwarfs likely does not follow the same functional form as around Sun-like stars. Furthermore, a comparison of occurrence rate at fixed $a$ may not be particularly informative, as these distances represent substantially different instellations when comparing between early and late M dwarfs.

\subsection{The influence of activity}

As stellar activity correlates with rotation rate and hence with radial-velocity uncertainty, choosing to restrict our analysis to inactive stars does not greatly affect our sensitivity to Jupiter-sized planets at the snow line: we would not have been able to detect such planets around most active stars due to the larger uncertainties.

However, two giant planets have been previously detected around 0.1--0.3M$_\odot$ stars: LHS 252 b, a 0.46M$_{\rm J}$ planet with a 204-day period \citep[][]{Morales2019}, and GJ 83.1 b, a 0.21M$_{\rm J}$ planet with a 771-day period \citep{Feng2020, Quirrenbach2022}. While both these stars are within 15 pc, we measure their H\textalpha\ emission to be in excess of our -1\AA\ cutoff and hence do not include them in our sample. In our four observations, both stars show statistically significant radial-velocity variations that are consistent with their published ephemerides; however, our observing cadence is too sparse to establish that activity has not contributed to the significance of these signals. It is intriguing that the only known giant planets around low-mass M dwarfs orbit active stars. Most active stars are rotationally broadened, meaning they have large radial-velocity uncertainties and it would be difficult to detect planets, even in the absence of activity-induced radial-velocity jitter. There are only 26 active, low-mass M dwarfs without close binary companions in our volume-complete sample that have $v$sin$i$ below 3 kms$^{-1}$, our detection threshold for line broadening. Two out of these 26 are already known to host a giant planet, and we measure $P(\chi^2) < 1\%$ for three out of the 26 (LHS 252, GJ 83.1, and also GJ 1224). In contrast, we do not detect any giant planets for our 200 inactive M dwarfs.

Is there a reason to expect giant planets to preferentially occur around active M dwarfs? A possible mechanism is the age--activity relation. Younger stars are more active \citep{Kiraga2007, Medina2022} and have higher metallicities on average. There is compelling evidence that giant planets in shorter period orbits are more common around metal-rich stars than metal-poor stars \citep{Santos2001, Fischer2005, Fulton2021, Rosenthal2022}, although the picture is less clear for M dwarfs \citep{Gaidos2014}. An inactive sample is therefore biased towards older, metal-poor, possibly planet-deficient stars. However, neither GJ 83.1 nor LHS 252 is particularly metal rich, with [Fe/H] estimates of -0.13 and 0.02 dex, respectively \citep{Newton2015}.

\subsection{A different occurrence distribution}
\label{sec:dist}
The planetary occurrence rate is often parameterized as a function of planetary mass and period \citep[e.g.,][]{Cumming2008}. As discussed above, \citet{Fulton2021} advocated for a broken power law parameterization for the giant planets of Sun-like stars. One might hypothesize that the giant planet occurrence function has the same shape around M-dwarfs as it does around Sun-like stars, with the entire distribution scaled down by some constant. However, our results suggest that this is not the case. If the activity of the known planet hosts from the previous section is a coincidence, the discrepancy in occurrence rates could instead be explained by the relatively low masses of LHS 252 b and GJ 83.1 b. LHS 252 b would fall in our sub-Jupiter bin, where we place a weaker 95\%-confidence upper limit of 4.4\%. For 0.21M$_{\rm J}$ planets like GJ 83.1 b, our sensitivity drops to only a handful of stars in our snow-line instellation bin ($N_{\rm eff}$ = 4). It may be that while Jupiter analogs are rare around low-mass M dwarfs, lower-mass giant planets still form, in a way that deviates from the uniform distribution in log($M_{\rm P}$sin$i$) found for the giant planets of Sun-like stars. Future surveys with higher sensitivity to 0.1--0.3M$_{\rm J}$ planets beyond the snow line will be necessary to explore this hypothesis.

However, lower-mass giant planets are unlikely to have a Jupiter-like impact on inner terrestrial planets, as Jupiter's migration profoundly shaped our inner solar system \citep{Walsh2011}. For Sun-like stars, a Jupiter-like mass is required to open a gap in the protoplanetary disk, or a Saturn-like mass in regions of low turbulent activity; smaller giant planets only partially open a gap in their disk \citep{Baruteau2013}, meaning they operate in a different regime of migration and lack Jupiter's capacity to suppress the inward flow of pebbles. While one might intuitively expect this gap-opening mass to decrease for lower-mass stars, the simulations of \citet{Sinclair2020} indicate that the threshold mass either increases or holds constant with decreasing stellar mass; those authors note that this behavior is the result of disks around lower-mass stars being geometrically thicker due to reduced gravity, yielding increased pressure forces. Lower-mass giant planets around M dwarfs will therefore not open Jupiter-like gaps.

Our results coupled with microlensing studies also hint at differing behavior with instellation. While the occurrence of giant planets peaks just beyond the snow line for Sun-like stars \citep{Fulton2021}, we find no giant planets at these instellations. Meanwhile, \citet{Poleski2021} analyzed 20 years of microlensing data to conclude that, on average, every microlensing star hosts a wide-orbit giant planet ($5 < a < 15$ au and $10^{-4} < M_{\rm p}/M_* < 0.033$); note that the microlensing lens distribution is dominated by M dwarfs \citep{Gould2010}. There have also been reported detections of $>1$M$_{\rm J}$ planets around low-mass M dwarfs at very wide orbital separations from direct imaging \citep{Gaidos2021} and disk studies \citep{Curone2022}. These results hint that giant planet occurrence may peak at lower instellations around low-mass M dwarfs as compared to Sun-like stars. Such a distribution could indicate a pathway for giant planet formation governed by processes unrelated to the water snow line, such as disk instability \citep{Boss2006, Mercer2020}. Other possible mechanisms to move giant planets to wide orbits include scattering by planet--planet interactions \citep[e.g.,][]{Rasio1996} or outward migration in resonance \citep[e.g.,][]{Crida2009}.

\begin{figure*}
    \centering
    \includegraphics[width=\textwidth]{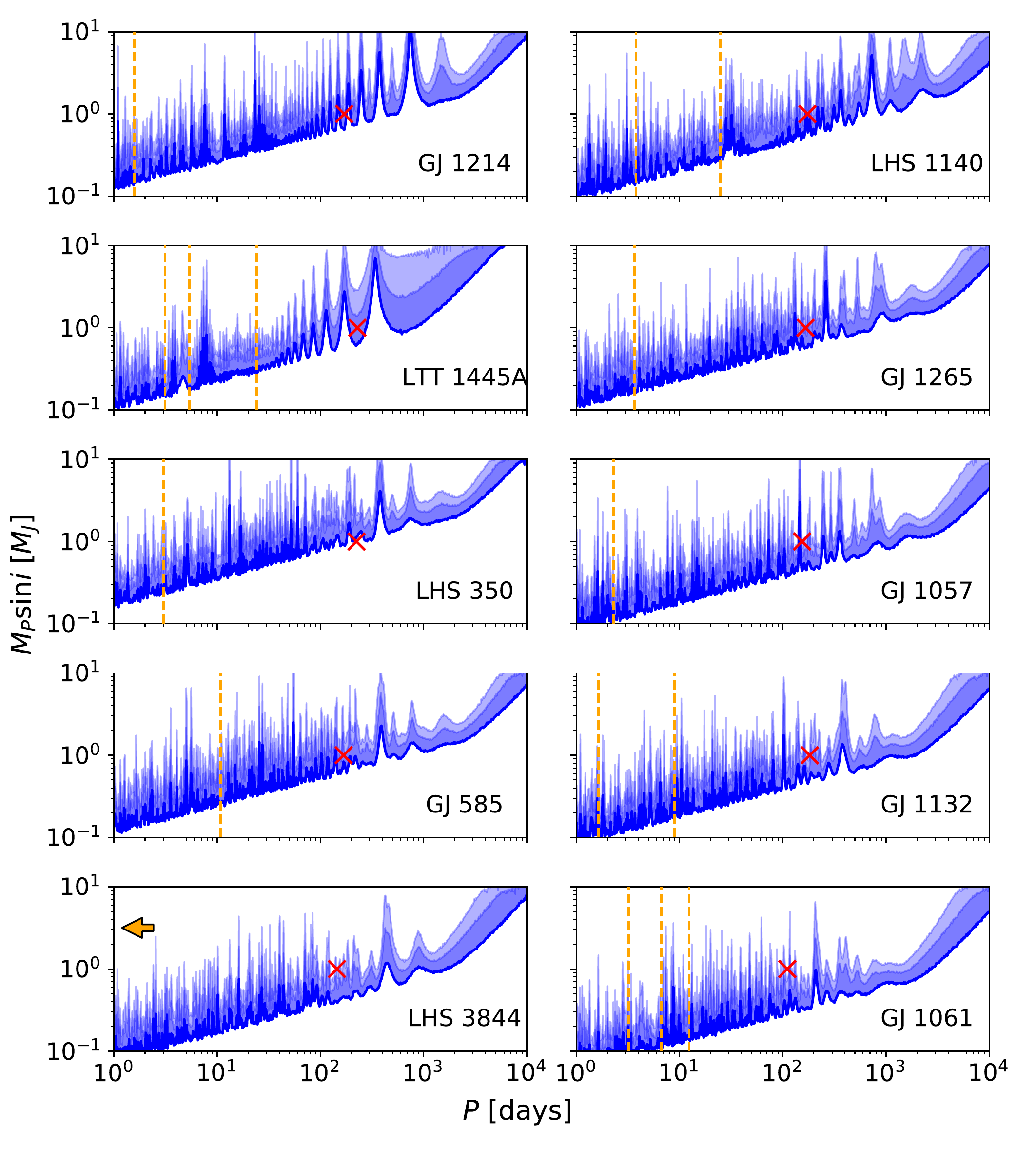}
    \vspace{-0.6cm}
    \caption{Our sensitivity to giant planets around the ten stars in our survey that are known to host a low-mass planet. The darkest blue line denotes a detection probability of 50\%. The contours illustrate where our survey places 84\% and 97.5\% constraints on the existence of a planet. A hypothetical planet with the instellation and mass of Jupiter is indicated with a red X. Dashed lines show the locations of the known small planets. LHS 3844 b is located off the left edge of the plot.}
    \label{fig:terr}
\end{figure*}

\subsection{Notable systems}

For Sun-like stars, \citet{Rosenthal2022} found that 41\% of stars with a close-in, small planet also hosted an outer giant, in contrast to a 17.6\% occurrence of outer giants overall. We are therefore particularly attentive to the sensitivity curves for the ten stars in our sample with published small planets: GJ 1214 \citep{Charbonneau2009}, GJ 1132 \citep{BertaThompson2015, Bonfils2018}, LHS 1140 \citep{Dittmann2017, Ment2019}, GJ 1265 \& LHS 350 \citep{Luque2018}, LHS 3844 \citep{Vanderspek2019}, LTT 1445A \citep{Winters2019, Winters2022, Lavie2022}, GJ 1061 \citep{Dreizler2020}, GJ 1057 \citep{Bauer2020}, and GJ 585 \citep{Harakawa2022}. While we do not have the radial-velocity sensitivity to recover these small planets, we can provide constraints on the existence of further-out giant planets to inform our understanding of exoplanetary system architectures.

In Figure~\ref{fig:terr}, we show our sensitivity to giant planets around each of these small-planet hosts. For all stars, we are likely to detect a planet that is Jupiter-like in mass and instellation, with our sensitivity varying from just under 50\% for LHS 350 to nearly 100\% for \hbox{GJ 1061}. As we only have four observations of each star, the sensitivity curves are bumpier than the smooth results of Figure~\ref{fig:occ}; for a given star, there are specific periods in which a planet could feasibly evade detection, although these hiding spots are averaged out when we consider the sample as a whole. For the purposes of Figure~\ref{fig:terr}, we have rerun our analysis with 1000 samples in mass and period to allow the nonhomogeneity of the sensitivity curves to be more readily appreciated. While we cannot definitively rule out the existence of a giant planet beyond the snow line for any individual system, Jupiter-analog companions to inner terrestrial planets -- i.e., planetary system architectures like our own -- cannot be commonplace for low-mass M dwarfs. If 41\% of these stars had a planet that was Jupiter-like in mass and instellation (a percentage motivated by the outer giant companion occurrence rate around Sun-like stars from \citealt{Rosenthal2022}), there is a 98\% chance we would have detected this planet around at least one of the ten stars. An abundance of lower-mass giant companions beyond the snow line (e.g., $M_{\rm P}$sin$i$ $\leq$ 0.3M$_{\rm J}$) is not ruled out by our observations: if all ten stars had a 0.3M$_{\rm J}$ planet at the Jovian instellation, there is a 78\% chance that we would have detected at least one, but this chance drops to 43\% when we lower the incidence to 41\%.

\subsection{Implications for terrestrial planets}

What does a lack of outer Jovians imply for the inner terrestrial planets in these systems (notably, the only terrestrial planets amenable to atmospheric study for at least the next decade)? While simulations on this topic are beyond the scope of this work, one can look to the large body of literature discussing the impact of Jupiter on the evolution of our own solar system. Models such as \citet{Walsh2011} suggest that the migration of Jupiter and Saturn led to the truncation of the planetesimal disk at 1 au, ultimately limiting the size of the terrestrial planets. Meanwhile, \citet{Izidoro2015} suggest that Jupiter acted as a dynamical barrier to the inward migration of gas giant cores, preventing them from migrating into the inner solar system from beyond the snow line and explaining why our solar system lacks a super-Earth (although note \citealt{Bryan2019}, who found that the presence of super-Earths correlates positively with the presence of an outer Jovian). Similarly, \citet{Bitsch2021} argue that a Jovian beyond the snow line can block the inward flow of water-rich pebbles, resulting in a drier inner stellar system. Systems without an outer Jovian may therefore have wetter (and potentially, larger) small planets; notably, \citet{Luque2022} recently identified a population of water worlds among the small planets that transit M dwarfs (although note \citealt{Ment2023}, who found that volatile-rich small planets are less common around mid-to-late M dwarfs). Giant planets are thought to also set the terrestrial water budget in our own solar system, with \citet{Raymond2017} arguing that Jupiter and Saturn were responsible for delivering Earth's small amount of water through planetesimal scattering.

\citet{Childs2022} note another implication of a lack of outer Jovians: without such a planet, a stellar system is unlikely to have an asteroid belt or a mechanism of delivering asteroids to inner terrestrial planets, and these asteroid impacts may be necessary for the origin of life \citep[e.g.,][]{Osinski2020}. Based on modelling from \citet{Childs2019}, \citet{Childs2022} further argue that lower-mass giant planets (specifically, they replaced Jupiter with a planet of mass 0.14M$_{\rm J}$) are unable to sustain suitable asteroid bombardment; more massive giant planets are needed. In the case of Earth, large asteroid impacts are thought to be responsible for the production of a reducing atmosphere \citep{Sinclair2020a}, which ultimately led to the emergence of prebiotic chemistry \citep{Benner2020}.

\section{Summary}
\label{sec:summary}

We present a null detection of giant planets in the volume-complete sample of 200 nearby, inactive 0.10--0.30M$_\odot$ M dwarfs. We place a 95\%-confidence upper limit of 1.5\% (68\%-confidence limit of 0.57\%) on the occurrence of $M_{\rm P}$sin$i > 1$M$_{\rm J}$ planets out to the water snow line ($T_{\rm eq} > 150$ K) around these low-mass M dwarfs. At the snow line ($100$ K $< T_{\rm eq} < 150$ K), we place 95\%-confidence upper limits of 1.5\%, 1.7\%, and 4.4\% (68\%-confidence limits of 0.58\%, 0.66\%, and 1.7\%) for 3M$_{\rm J} < M_{\rm P}$sin$i < 10$M$_{\rm J}$, 0.8M$_{\rm J} < M_{\rm P}$sin$i < 3$M$_{\rm J}$, and 0.3M$_{\rm J} < M_{\rm P}$sin$i < 0.8$M$_{\rm J}$ giant planets. More granular constraints are given in Figure~\ref{fig:occ}.

Planets that are Jupiter-like in mass and instellation are rare around low-mass M dwarfs, consistent with expectations from core accretion theory. Compared with previous radial-velocity and microlensing studies that consider broader distributions of M dwarfs with higher mean stellar masses, our results are consistent with a decrease in giant planet occurrence with decreasing M-dwarf mass, although direct comparison between surveys is complicated by the limited bins in which occurrence rates have been published and the fact that a given bin in $P$ or $a$ around a more massive star corresponds to a much lower instellation around a low-mass M dwarf. In addition, the picture of giant planet occurrence from microlensing is still unclear. If \citet{Poleski2021} are correct in their assertion that every microlensing star has a wide-orbit giant planet, our results imply that the distribution of giant planets around low-mass M dwarfs must differ dramatically from more massive stars, whose giant planets are more prevalent near the water snow line than on wide orbits.

A lack of Jupiter analogs around low-mass M dwarfs has profound impacts for the formation and evolution of habitable-zone terrestrial planets, from their sizes and compositions, to their dynamical environment, to their volatile and refractory budgets, and perhaps to their capacity for life. While terrestrial planets that transit low-mass M dwarfs are promising targets for atmospheric characterization, these worlds will have evolved in a vastly different environment to our own.

\section*{Acknowledgements}

We thank the TRES, CHIRON, and MAROON-X teams for their support, including Jacob Bean, Allyson Bieryla, Lars Buchhave, Pascal Fortin, Todd Henry, Hodari James, Wei-Chun Jao, David Kasper, Leonardo Paredes, Samuel Quinn, Andreas Seifahrt, Andrew Szentgyorgyi, and Andrei Tokovinin, as well as Amber Medina for advice in measuring H\textalpha\ equivalent widths.

EP is supported in part by a Natural Sciences and Engineering Research Council of Canada (NSERC) Postgraduate Scholarship. This work is made possible by a grant from the John Templeton Foundation. The opinions expressed in this publication are those of the authors and do not necessarily reflect the views of the John Templeton Foundation.

This work is based on observations obtained at the international Gemini Observatory, a program of NSF’s NOIRLab, which is managed by the Association of Universities for Research in Astronomy (AURA) under a cooperative agreement with the National Science Foundation. on behalf of the Gemini Observatory partnership: the National Science Foundation (United States), National Research Council (Canada), Agencia Nacional de Investigaci\'{o}n y Desarrollo (Chile), Ministerio de Ciencia, Tecnolog\'{i}a e Innovaci\'{o}n (Argentina), Minist\'{e}rio da Ci\^{e}ncia, Tecnologia, Inova\c{c}\~{o}es e Comunica\c{c}\~{o}es (Brazil), and Korea Astronomy and Space Science Institute (Republic of Korea). These observations were obtained under program GN-2022A-FT-205.

This work has made use of data from the European Space Agency (ESA) mission
{\it Gaia} (\url{https://www.cosmos.esa.int/gaia}), processed by the {\it Gaia}
Data Processing and Analysis Consortium (DPAC,
\url{https://www.cosmos.esa.int/web/gaia/dpac/consortium}). Funding for the DPAC
has been provided by national institutions, in particular the institutions
participating in the {\it Gaia} Multilateral Agreement.

This publication makes use of data products from the Two Micron All Sky Survey, which is a joint project of the University of Massachusetts and the Infrared Processing and Analysis Center/California Institute of Technology, funded by the National Aeronautics and Space Administration and the National Science Foundation.

This research has made use of the SIMBAD database,
operated at CDS, Strasbourg, France.

\facilities{CTIO:1.5m (CHIRON), FLWO:1.5m (TRES), Gemini:Gillett (MAROON-X)}
\software{\texttt{astropy} \citep{Astropy2013, Astropy2018}, \texttt{emcee} \citep{ForemanMackey2013}, \texttt{matplotlib} \citep{Hunter2007}, \texttt{numba} \citep{Lam2015}, \texttt{numpy} \citep{Harris2020}, \texttt{radvel} \citep{Fulton2018}, \texttt{scipy} \citep{Scipy2020}}


\bibliography{pass2023}{}
\bibliographystyle{aa_url}



\end{document}